# *healthAIChain*: Improving security and safety using Blockchain Technology applications in AI-based healthcare systems

Naresh Kshetri
*Dept. of Acct. & Technology*
*Emporia State University*
Emporia, Kansas, USA
NKshetri@emporia.edu

James Hutson
*Dept. of Art History & Culture*
*Lindenwood University*
St. Charles, Missouri, USA
JHutson@lindenwood.edu

Revathy G
*Dept. of Computer Science*
*SRC, SASTRA Deemed University*
Thanjavur, Tamilnadu, India
RevathyJayabaskar@gmail.com

*Abstract* — Blockchain as a digital ledger for keeping records of digital transactions and other information, it is secure and decentralized technology. The globally growing number of digital population every day possesses a significant threat to online data including the medical and patients' data. After bitcoin, blockchain technology has emerged into a general-purpose technology with applications in medical industries and healthcare. Blockchain can promote highly configurable openness while retaining the highest security standards for critical data of medical patients. Referred to as distributed record keeping for healthcare systems which makes digital assets unalterable and transparent via a cryptographic hash and decentralized network. The study delves into the security and safety improvement associated with implementing blockchain in AI-based healthcare systems. Blockchain-enabled AI tackles the existing issues related to security, performance efficiencies, and safety in healthcare systems. We have also examined the Artificial Intelligence in healthcare and medical industry, potential areas, open questions concerning the blockchain in healthcare systems. Finally, the article proposed an AI-based healthcare blockchain model (healthAIChain) to improve patients' data and security.

*Keywords* — applications, artificial intelligence, blockchain, healthcare, patients' data, safety, security

## I. INTRODUCTION

The merger of Artificial Intelligence (AI) into healthcare systems offers an avenue for enhanced service provision and potential cost reduction. Notable technology conglomerates, such as Google and Microsoft, are making significant strides towards the development and refinement of AI, demonstrating a promising future for AI's role in healthcare. Leveraging the capabilities of AI could escort to significant enhancements in patient management and the minimization of medical errors [1]. Further, the implementation of the technology has the potential to augment the efficiency of physicians and elevate the quality of medical services [2]. Moreover, AI emerges as a transformative force within the healthcare sector, possessing the capacity to mitigate the existing human resources crisis. However, the deployment of AI raises salient ethical considerations that require addressing. For instance, a potential imbalance could emerge between those who adopt AI and those who refrain from its usage [3]. Yet, preparation for such shifts in the landscape remains plausible.

Recent research has underscored the necessity for investments in AI specifically for public health practice. Yet, to achieve significant benefits, such investments must be accompanied by soared access to high-quality data and comprehensive education on the restrictions of AI. Additionally, the establishment of rigorous regulation is crucial for its safe and effective implementation [4]. The rapid growth of AI in medicine showcases its potential for widespread impact, including clinical settings, data science, and policymaking. Moreover, it bears the promise of reducing healthcare inequality between developing and developed countries [5]. However, patient apprehensions concerning the safety, patient choice, healthcare fees, data-source unfairness, and data security of AI in healthcare persist. Addressing these concerns is critical for the long-term success of AI in healthcare [6]. This research will delve into the potential of employing blockchain technology to discourse some of these concerns, particularly those related to data security and safety.

Given the transformative potential of AI in healthcare, alongside the emerging challenges, there is a compelling need for new frameworks to ensure security, safety, and efficacy. As a result, this paper will propose the implementation of a blockchain-enabled AI model, referred to as *HealthAIChain*, that has the potential to tackle existing issues related to security and performance efficiencies in healthcare systems. Grounded in the robustness and transparency of blockchain technology, *HealthAIChain* seeks to provide unalterable and secure record keeping for healthcare systems, thereby enhancing the safety of critical patient data. This research aims to illuminate the vast possibilities at the intersection of AI and blockchain technology, ultimately presenting a novel model that could serve to catalyze significant advancements within the domain of digital healthcare. The ensuing discourse will shed light on the underlying mechanisms of blockchain technology, its potential applications within AI-based healthcare systems, and the key components of the proposed *healthAIChain* model.

## II. RELATED WORK

The body of literature exploring the intersection of blockchain technology and healthcare has grown significantly in recent years. A variety of studies have examined the opportunities, challenges, and potential solutions offered by blockchain within this vital sector. For instance, an analysis by McGhin et al. [7] reveals that many current blockchain experiments do not address unique requirements posed by healthcare applications. Furthermore, Abdu and Wang [8] underscore the importance of quality node verification and addressing anomaly access of information from blocks within the healthcare setting. Highlighting the potential of blockchain, Prokofieva and Miah [9] recognize blockchain-based approaches as transformative for healthcare information dissemination.

A systematic review by Odeh et al. [10] underscores the significance of principled awareness among healthcare professionals. Simultaneously, Hasselgren et al. [11] point to the exponential increase in endeavors to utilize blockchain technology in the health domain. On the other hand, Javaid et al. [12] spotlight the significance of blockchain in mitigating the fear of data manipulation within healthcare. Notably, several researchers highlight the potential for blockchain to enhance data efficiency and security. For instance, Yaqoob et al. [13] propose the creation of applications for managing and sharing secure, transparent, and immutable audit trails. Gökalp et al.[14] maintain that blockchain adoption could secure communications among stakeholders, a sentiment echoed by Huifen. Xu et al. [15], who claim that blockchain can improve accessibility and security of patient information.

Blockchain's disruptive potential has not gone unnoticed. Reddy and Madhushree [16] argue that blockchain can enhance data efficiency, data access flexibility, interconnection, and transparency in the healthcare sector. Concurrently, Chukwu and Garg [17] note blockchain's promise in addressing electronic health record (EHR) challenges. Yet, despite the burgeoning promise of blockchain, there are recognized challenges and barriers. Umrao et al. [18] point out the difficulties of incorporating blockchain into IoT systems within the healthcare sector, while Agbo and Mahmoud [19] discuss hurdles related to interoperability, security-privacy, scalability-speed, and stakeholder engagement.

However, the proliferation of research in the area suggests a burgeoning confidence in the transformative potential of blockchain in healthcare. Through comprehensive review and analysis, these studies collectively contribute to the robust discourse on the potential, challenges, and opportunities of merging blockchain technology within the healthcare sector. Building upon the foundational work of these researchers, further exploration of the intersections between blockchain technology and healthcare can be taken in several directions. To address concerns around the unique requirements of healthcare applications as raised by McGhin et al. [7], further research could focus on creating blockchain solutions specifically tailored for the healthcare sector.

Moreover, the issues regarding the quality of nodes and anomaly access of information raised by Abdu and Wang [8] warrant detailed investigation. Research could be aimed at enhancing the reliability and integrity of nodes in the blockchain network, especially within the context of sensitive health data, and at developing novel anomaly detection techniques to protect patient information. Given the substantial potential for blockchain-based approaches in healthcare information dissemination as noted by Prokofieva and Miah [20], future investigations could focus on the development of scalable and secure platforms for sharing healthcare data. Furthermore, studies could delve into blockchain's potential for enhancing patient privacy and security while facilitating data sharing.

The observation by Hasselgren et al. [21] of the exponential increase in efforts to use blockchain technology within the health domain underlines the importance of understanding the implications of this rapid expansion. Researchers could assess the implications of these endeavors on the broader healthcare sector, examining the ethical, legal, and social dimensions of widespread blockchain adoption. Additionally, studies could also address the challenges identified by researchers such as Umrao et al. [22] and Agbo and Mahmoud [19]. Investigations could be undertaken to devise strategies for effectively integrating blockchain with IoT systems in healthcare, and overcoming the hurdles related to interoperability, privacy, scalability, and stakeholder engagement. Overall, while significant progress has been made in understanding the potential applications of blockchain technology in healthcare, these contributions also reveal an array of challenges and research avenues that remain to be fully explored. Future investigations should continue to build upon these initial findings to further our acknowledging of how blockchain can be effectively and responsibly implemented within the healthcare sector.

## III. HEALTHCARE USING AI

The potential part of AI in healthcare has become a research focal point, offering the promise of improving various aspects of care. According to Shaheen [23], AI-assisted clinical trials can process massive data volumes, producing highly accurate results. Furthermore, Väänänen et al. [24] highlight the dual advantages of cost reduction and improved health outcomes as AI's benefits in healthcare, while Puaschunder and Feierabend [25] emphasize AI's advanced capabilities in information gathering, processing, and delivering well-defined outputs to end-users. Meanwhile, Davenport and Kalakota [26] propose that AI can execute healthcare tasks as well or better than humans. This assertion gains support from Secinaro et al. [27], who note that AI can aid physicians with diagnoses, disease spread predictions, and treatment path customization.

The advancement of AI methods, including machine learning and deep learning, have improved the efficiency of diagnosis and prognosis, as observed by Houfani et al. [28]. An array of applications for AI in healthcare is also acknowledged by Sharma and Kumar [29], further underscoring the technology's potential. Whig et al. [30] emphasize AI's role as a critical time and effort-saving tool for healthcare firms. Palmer and Short [31] sees AI as the next evolutionary step in computer use, leveraging data to enhance user efficiency. Roy and Jamwal [32] provide examples from India, where AI assists in delivering diagnostic and prescriptive solutions in leading hospitals. The field of AI in medicine, particularly cardiology and brain science, is rapidly evolving.

Matheny et al. [33] suggest AI's potential role as complementary to human cognition in delivering personalized health care. The AI's promise, according to Aung et al. [34], lies in its potential to enhance healthcare but also necessitates careful governance akin to the regulation of physician conduct. Exploring further, Shenoy et al. [35] highlight how AI-based technologies are expanding into uncharted areas of healthcare. The enhancement of processes related to speed, cost, capacity, quality, and consistency is attainable through AI. Sunarti et al. [36] stress the need for AI implementation in health service management efficiency and medical decision-making.

On the other hand, Chen and Décary [37] provide an essential guide for health leaders, suggesting that AI can improve health service efficiency, safety, and access, while Maddox, Rumsfeld, and Payne [38] see AI as a revolutionary force for patients and populations in healthcare. However, Jadhav and Boraste [39] sound a note of caution, arguing that although AI algorithms already surpass human experts today, the use of AI in healthcare hitches up ethical questions. Similarly, Boehlen et al. [40] argue that AI-based clinical tools can challenge commonly held values and ethical principles.

The theme of augmented the decision-making abilities of clinicians has also risen as a central theme. As Silvana Secinaro et al. [41] observe, AI can bolster physicians in making diagnoses, predicting disease spread, and customizing treatment paths. Enhanced with machine learning and deep learning techniques, AI can analyze intricate patient data, detect nuanced patterns, and predict disease trajectories, thereby enabling proactive and personalized care. Complementing this perspective, T. Davenport, and Ravi Kalakota [42] posit that AI can accomplish healthcare tasks equally or more effectively than humans. AI's role in diagnosing diseases, developing treatment plans, and providing personalized care demonstrates its potential to significantly impact patient outcomes and healthcare delivery.

Efficiency and cost reduction are pivotal areas where AI can significantly contribute. Väänänen et al. [24], as noted, argue that AI implementation in healthcare can achieve considerable cost reductions while improving health outcomes. Through optimization of resource allocation and reduction of operational inefficiencies, AI has the potential to deliver high-quality care while simultaneously curtailing expenses. At the same time, Sharma, and Kumar [43] note the indispensability of AI for healthcare firms in saving time and effort on specific tasks. The automation of routine tasks can free healthcare professionals to focus on complex patient care tasks, enhancing service delivery and potentially patient outcomes.

The role of AI in handling vast amounts of healthcare data is also noteworthy. The work of Shaheen [23] underscores the potential of AI in clinical trials, particularly its capacity to manage large data volumes and generate precise results. Such capabilities can speed up the process of clinical trials, leading to faster discoveries and deployment of treatments. Puaschunder and Feierabend [44] emphasize AI's advanced abilities to gather, process, and deliver well-defined outputs. These capabilities can equip clinicians with comprehensive and actionable patient data, enabling more informed decision-making and personalized care strategies.

The ethical and governance aspects of AI in healthcare constitute a significant theme in the literature. Aung et al. [34] assert that the introduction of AI in healthcare needs meticulous governance, akin to the conduct regulation of physicians. These regulatory and ethical considerations are crucial for the responsible and equitable use of AI in healthcare. Therefore, AI has the potential to reorganize healthcare, influencing clinical decision-making, optimizing efficiencies, enhancing clinical trials, and reshaping patient care. However, the ethical and governance aspects warrant careful consideration to ensure the responsible use of AI in the best interests of patients and society.

## IV. Improving Security and Safety

The concerns about security and safety as well as transparency and trust is the central theme for healthcare systems as the data dealing with healthcare are real-time critical data. Blockchain technology decrease the possibility of unsecured transactions because of decentralization and non-intermediaries to monitor the transactions between participating patients. The consensus mechanism of blockchain technology besides other prime characteristics of decentralization, immutability, traceability, and privacy [45] can be useful to enhance the overall security of critical systems like healthcare and the medical industry.

Transparency and trust has a immediate bond to corruption and citizen's satisfaction with public services including healthcare. Security and safety improvement in healthcare systems are directly proportional to improved transparency and citizen's trust by goals to uplift the living standards of publics including patients [46]. As a promising technology, blockchain technology has the amplitude to accredit transparency and erect citizen's trust in resident service delivery while preserving an ample level of privacy and security. In variation to client-server architecture, blockchain technology is rooted on P2P architecture, where power doesn't pause on a lone system, and can be a promising tool to enhance security and medical safety.

Another important concern for improving safety and security is the swift rise of systems and the internet which heightened the worldwide upswing of online users and connected devices. The everyday work we are doing online and the documents we are splitting online, can be easily retrieved by an awful actor when a computer is breached. After the COVID-19 epidemic, many of the cyberattacks use the coronavirus motif directly or indirectly also in the healthcare sector to gain the user's trust [47]. Cyber threats have beginnings in the banking, business belt, and military area as well including the healthcare industry that uses many AI-controlled devices for the real time diagnostic things.

The forthcoming new internet infrastructures will certainly be at significant risk to improve security and safety if the issues are not handled effectively.

The history of internet crime and cyberattacks in pre and post COVID era has affected the survival of every person directly or indirectly in concern for improving security and safety [48]. Major branches of online crime include URL redirection, cyberstalking, online scams, credential phishing including identity theft that impacted the healthcare industry a lot during the coronavirus phase. The boundless use of social media, work from home, reduced traffic in communal places, rising number of web connected devices with internet users etc. is steering to online crime, which is also a primary cause of cyberattacks. All person in the web cyber space is distressed about their own security, and people lack primary cyber ethics before becoming involved in the defense of networks that lead to another hurdle for improving safety and security in the cyber domain.

Emergence of innovative technologies like Artificial intelligence, augmented reality, virtual reality, metaverse, blockchain technology, and natural language processing are modifying how businesses communicate with customers by providing personalized feels [49]. The remarkable security and privacy disputes emerging from these new technologies and improvements with aid to data mining, customer data usage, and virtual surroundings is also gaining popularity in the healthcare and medical industry. Businesses including the medical, healthcare industry need to cope to the growing digital landscape and are essential to stay competitive when it comes to improvement of security and safety of business customers including health patients. The rising authority of social media platforms, influencer medical marketing, and voice hunt gadget over the web is changing the future of the digital marketing and artificial intelligence world.

## V. BCT for AI-Based Healthcare

One of the most cutting-edge technologies accessible today to guarantee the protection of consumers' sensitive or secret data is blockchain technology. Many applications, including artificial intelligence, supply chains, cloud computing, the healthcare industry, and countless, rely heavily on blockchain technology. It enables the healthcare industry to take advantage of its numerous cutting-edge characteristics, including privacy, security, decentralization, and secrecy. Also, the Internet of Things (IoT) devices connect with healthcare systems, and the application program for the healthcare sector also interfaces with the IT sector. The healthcare industry has been profoundly impacted by blockchain-based IoT systems because they improve security, privacy, transparency, and efficiency while presenting improved economic options. However, there are other security and privacy issues that traditional healthcare systems must deal with, including phishing, identity theft, and masquerades [50].

Many diseases are causing the world to face numerous healthcare issues. Blockchain is a promising and secure technology that is being used to develop new, creative concepts as well as solutions in many industries, including banking, supply chains, agriculture, and the healthcare system, but it is not the answer to all problems. With the help of hospitals, diagnostic labs, pharmacy companies, and doctors, a blockchain technology networking is employed in the healthcare domain to keep and replace patient data. Applications based on the blockchain technology can precisely identify serious errors in diagnosis, even those that are harmful. Hence, blockchain technology can magnify the efficacy, security, and transparency of exchanging medical data throughout the healthcare system [51].

The hottest technology right now is called the metaverse, which is attracting interest from both academics and business. For increased usability, many parties involved are thinking about integrating their current apps into the metaverse ecosystem. The metaverse is gradually being utilized by the healthcare sector to raise living standards and improve service quality. In this article, we specifically discuss the possibility for electronic anti-aging healthcare in a metaverse setting. Demonstrate how the metaverse environment may be castoff to improve healthcare service quality and lengthen patient lives via more safe processes, such as chronic illness management, physical fitness, and mental health guidance. Healthcare practitioners can accelerate the anti-aging process by utilizing these technologies to enhance patient outcomes, lower healthcare prices, and provide innovative healthcare experiences for a superior life. A secure and open ecosystem for healthcare data can be created using blockchain, while AI can be castoff to evaluate infinite amounts of medical data and develop individualized treatment regimens [52].

The protection of patients' medical information is made possible by contemporary technologies like edge computing, blockchain, and machine learning. In light of latter growth in blockchain technology, such as the erasure of contaminating assaults, helping farmers such as sugarcane industry, and the easy supply of total transparency and data integrity over the decentralized system. Employing the model at the edges safeguards the cloud from threats by down-sizing details from its gateway with few computing time and processing potential because federated AI works with lesser datasets [53][54].

Automation and digitization have always had a significant strike on the healthcare industry. It holds all cutting-edge and novel technologies. The leap of the metaverse with AI and blockchain, a new technology in the digital realm, has freshly been seen all over the world. With an immersive involvement, the metaverse has enormous potential to offer a wide range of health services to patients and health professionals. The integration of blockchain technology and artificial intelligence in the metaverse & finance, business to deliver more effective, efficient, and secure healthcare, medical facilities in virtual space [55] [56].

## VI. Proposed Model - HealthAIChain Model

We have proposed a healthAIChain model with an aim to improve patient's data and security. There are various components of the proposed model, healthAIChain that integrates blockchain technology with AI-based healthcare systems which are described below. We have also shown and described the architecture of the proposed system in the figure below apart from the healthAIChain algorithm.

a. Patient or patient's data - the patient is the primary user and bearer of the model or application, his/her login details are

documented in the blockchain network in cryptographically secured form.

<u>b. Wallet of patient</u> - includes the patient's app (also known as user / patient wallet) with data for login etc. that verifies the login success for the patient or user.

<u>c. Operational framework</u> - is the interoperability framework after the patient's wallet that leads and allows public services and patient networks.

<u>d. Services for patient</u> - patient services or the public services with public key available only after the successful login of the patient in the model.

<u>e. BCT network (with consensus node)</u> - the central network of the proposed model that includes nodes, smart contracts, and identifies the model.

<u>f. Header & transactions (encrypted)</u> - the transactions and header (private) for the patient record after the patient session is permitted to the user.

<u>g. Patient record manager (database)</u> - the open ledger for trust of information for patient data, trusted room among patients for storing and retrieving patient data.

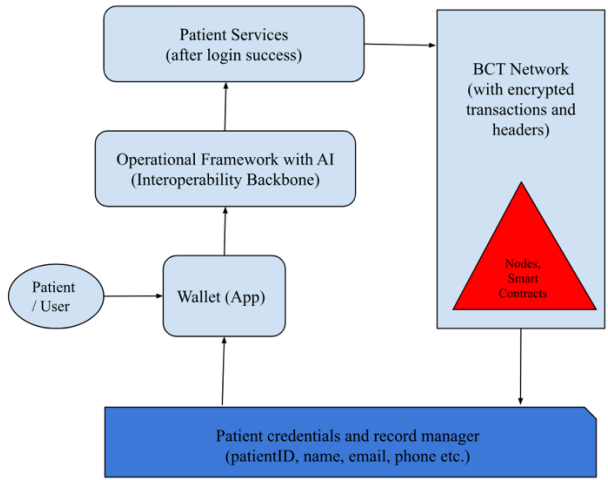

Fig. 1. Proposed model for healthAIChain to improve patients' data and security, i.e., AI - based healthcare system using blockchain technology

## VII. CONCLUSION AND FUTURE SCOPE

The improvement of security and safety in healthcare systems is very urgent due to the increasing volume of online data and number of internet population every year. The merger of AI and blockchain technology can be a perfect match with applications in medical industries, patient's data, and healthcare. The emergence of AI as a transformative force within the healthcare sector and emergence of blockchain technology to validate transparency and citizen's trust in public service delivery can be benefitted in healthcare systems. Blockchain-enabled AI tackles the existing issues related to security, performance, immutability, and safety with the proposed HealthAIChain model to shield patient's data. The pseudocode of the proposed model helps to add new patient details with a public ID that can be used for patient services after successful login with a blockchain network including nodes.

Future research should include a more in-depth analysis, training the model to address challenges related to government and private healthcare sectors for including blockchain technology. Blockchain is not only a database technology and a distributed ledger but also a computing means with several nodes that can benefit both patients and doctors. The automation process with the assistance of artificial intelligence combined blockchain technology can compensate healthcare systems in several other ways than patient's data safety and security.